# Translation Invariant Bipolarons and Charge Density Waves in High-Temperature Superconductors

**Victor D. Lakhno[1*]**

[1]Keldysh Institute of Applied Mathematics of Russian Academy of Sciences, Moscow, Russia

**\* Correspondence:**
Victor Lakhno
lak@impb.ru

**Abstract**

A correlation is established between the theories of superconductivity based on the concept of charge density waves (CDW) and the translation invariant (TI) bipolaron theory. It is shown that CDW are originated from TI-bipolaron states in the pseudogap phase due to Kohn anomaly and form a pair density wave (PDW) for wave vectors corresponding to nesting. Emerging in the pseudogap phase, CDW coexist with superconductivity at temperatures below that of superconducting transition while their wave amplitudes decrease as a Bose condensate is formed from TI-bipolarons, vanishing at zero temperature.

## 1      Introduction

Presently there is no agreement about the microscopic nature of the high-temperature superconductivity (HTSC). At the same time there are phenomenological models such as a Ginzburg–Landau model [1], a model of charge density waves (CDW) or pair density waves (PDW) (the recent review on the theory and experiment with CDW/PDW in high temperature superconductors, ultra-could atomic gases and mesoscopic devices was published by Agterberg et al. [2]) and a model of spin density waves which enable one to describe numerous HTSC experiments [3]. These models are silent on the nature of paired states taking part in SC. In author's works [4-7] by paired states are meant translation invariant (TI) bipolaron states formed by a strong electron-phonon interaction similar to Cooper pairs. (For the review of earlier pioneering works devoted to superconductivity based on the theory of small radius bipolarons see [8]). TI-bipolarons are plane waves with a small correlation length capable of forming a Bose-Einstein condensate with high transition temperature which possesses SC properties. A correlation between the Bardeen-Cooper-Schrieffer (BCS) theory [9] and Ginzburg–Landau theory was established by Gor'kov [10]. The aim of this work is to establish a correlation between the TI-bipolaron theory of SC and CDW (PDW).

## 2      Results

### 2.1   General relations for the spectrum of a moving TI bipolaron

TI-bipolarons are formed at a temperature $T^*$ which is much higher than that of a SC transition $T_c$ (refs. [4-7]). For $T_c < T < T^*$ and in the absence of a Fermi surface with a sharp boundary, an ensemble of TI-bipolarons would be an ideal gas whose particles would have a spectrum $s = \{v_k^2(\vec{\mathcal{P}})\}$, determined by the dispersion equation [4-7]:

$$1 = \frac{2}{3}\sum_k \frac{k^2|f_k|^2 \omega_k}{s - \omega_k^2}, \tag{1}$$

$$\omega_k = \omega_0(\vec{k}, \vec{\mathcal{P}}) - \frac{\vec{k}\vec{\mathcal{P}}}{M} + \frac{k^2}{2M} - \frac{\vec{k}}{M}\sum_{k'}\vec{k}|f_{k'}|^2,$$

$f_k = f_k(\vec{k}, \vec{\mathcal{P}})$ are parameters determining the ground state energy $E_{bp}(\vec{\mathcal{P}})$ of a TI-bipolaron, $\vec{\mathcal{P}}$ is the total momentum of a TI bipolaron, $M = 2m$, where $m$ is the mass of a band electron (hole), $\omega_0(\vec{k}, \vec{\mathcal{P}})$ is the phonon frequency in an electron gas around TI-bipolarons.

In describing Kohn anomaly, one usually proceeds from Fröhlich Hamiltonian of the form [11]:

$$H = \sum_p \varepsilon(\vec{p}) c_{\vec{p}}^+ c_{\vec{p}} + \sum_{\vec{p}} \hbar\tilde{\omega}(k) b_{\vec{k}}^+ b_{\vec{k}} + \sum_{\vec{p}\vec{k}} g(\vec{k}) c_{\vec{p}+\vec{k}}^+ c_{\vec{p}} (b_{-\vec{k}}^+ + b_{\vec{k}}), \tag{2}$$

where the first term corresponds to a free electron gas; $c_p^+, c_p$ are operators of the birth and annihilation of an electron with energy $\varepsilon(\vec{p})$. The second term corresponds to Hamiltonian of the lattice; $b_{\vec{k}}^+, b_{\vec{k}}$ are operators of the birth and annihilation of lattice oscillations with energy $\tilde{\omega}(k)$. The third term describes interactions of electrons with the lattice; $g(k)$ is the matrix element of the interaction.

Renormalization of phonon frequencies corresponding to (2) is determined by the expression [11]:

$$\omega^2(\vec{k}) = \tilde{\omega}(\vec{k})^2 + 2\tilde{\omega}(\vec{k})|g(\vec{k})|^2 Re[\chi(\vec{k})], \tag{3}$$

where $\omega(\vec{k})$ are phonons renormalized by an interaction with an electron gas whose polarizability is determined by $\chi(\vec{k})$.

Kohn anomaly describes vanishing of renormalized phonon modes $\omega(k)$ for $k = \mathcal{P}_{CDW}$.

In the TI bipolaron theory of SC [4-7] bipolarons are believed to be immersed into an electron gas. The properties of such bipolarons are also described by Fröhlich Hamiltonian of the form (2), but with a field of already renormalized phonons with energies $\omega(k)$ and, accordingly, a matrix element of the interaction $V(\vec{k})$ instead of $g(\vec{k})$.

It should be noted that spectral equation (1) is independent of the form of $V(\vec{k})$.

The experimental evidence of the occurrence of renormalized phonons with zero energy in layered HTSC cuprates is the absence of a gap in the nodal direction in these materials which, by definition, in the TI bipolaron theory of SC is the phonon frequency.

The wave function of a TI-bipolaron with the wave vector $\vec{\mathcal{P}}$ will have the form:

$$|\Psi(\vec{\mathcal{P}})\rangle_{bp} = e^{i\vec{\mathcal{P}}\vec{R}}|\Psi(0)\rangle_{bp} \tag{4}$$

where $\vec{R}$ are coordinates of the bipolaron mass center. The explicit form of the wave function $|\Psi(0)\rangle_{bp}$ with zero momentum is well established [12]. An expression for $\vec{\mathcal{P}}$ can be derived by calculating the mathematical expectation of the operator of the total momentum $\hat{\mathcal{P}}$:

$$\vec{\mathcal{P}} = \langle \Psi(\vec{\mathcal{P}})|\hat{\mathcal{P}}|\Psi(\vec{\mathcal{P}})\rangle = M\vec{u} + \sum_k \vec{k}|f_k|^2, \tag{5}$$

where $\vec{u}$ is the velocity of a TI-bipolaron. Assuming that $\vec{\mathcal{P}} = M_{bp}\vec{u}$, where $M_{bp}$ is the bipolaron mass, we can rewrite (5) and express $M_{bp}$ as:

$$M_{bp} = \frac{M}{1-\eta}, \qquad \eta = \frac{\vec{\mathcal{P}}}{\mathcal{P}^2}\sum_k \vec{k}|f_k|^2. \tag{6}$$

With the use of (6) expression for $\omega_k$ from (1) can be rewritten as:

$$\omega_k = \omega_0(\vec{k},\vec{\mathcal{P}}) + \frac{k^2}{2M} - \frac{\vec{k}\vec{\mathcal{P}}}{M_{bp}}. \tag{7}$$

It follows from (6) that in the case of a weak and intermediate coupling (when TI-bipolaron states are metastable for $\vec{\mathcal{P}} = 0$), the exact form of $f_k(\vec{k},\vec{\mathcal{P}})$ is known [13] and in the case of bipolaron equals to:

$$f_k(\vec{k},\vec{\mathcal{P}}) = -\frac{V_k}{\omega_0(\vec{k},\vec{\mathcal{P}}) - \frac{\vec{k}\vec{\mathcal{P}}}{2M}(1-\eta) + \frac{k^2}{2M}},$$

where $V_k$ is the electron-phonon matrix element and the expression for the TI-bipolaron effective mass will have a simple form: $M_{bp} = M(1 + \alpha/6)$, where $\alpha$ is a constant of electron-phonon interaction, that is the mass $M_{bp}$ is equal to the sum of masses of individual polarons. For large $\alpha$, good approximations for $f_k$ are available only for $\vec{\mathcal{P}} = 0$. For this reason, the calculation of the effective mass of a TI-bipolaron is rather difficult.

## 2.2   TI-bipolarons and charge density waves

Subject to availability of a Fermi surface with a sharp boundary, the TI-bipolaron gas under consideration will have some peculiarities. Thus, if there are rather large fragments on this surface which can be superimposed by transferring one of them onto vector $\vec{\mathcal{P}}$, then if the size of these fragments is rather large, the coupling between them will be rather strong which will lead to Peierls deformation of the lattice in the direction of this nesting. A loss of energy associated with the lattice deformation will be compensated by a gain in the energy of a bipolaron gas which forms a charge density wave with the wave vector $\vec{\mathcal{P}} = \vec{\mathcal{P}}_{CDW}$. The fact of a gain in the energy of a TI bipolaron follows from equation (1), whose solution leads to the spectrum of a TI-bipolaron $E_k(\vec{\mathcal{P}})$, where:

$$E_k(\vec{\mathcal{P}}) = \begin{cases} E_{bp}(\vec{\mathcal{P}}), & k = 0 \\ E_{bp}(\vec{\mathcal{P}}) + \omega_0(\vec{\mathcal{P}},\vec{k}) + \frac{k^2}{2M} - \frac{\vec{k}\vec{\mathcal{P}}}{M_{bp}}, & k \neq 0 \end{cases}, \tag{8}$$

i.e.: $E_k = E_{bp} + \omega_k$, and $v_k = \omega_k$ for $k \neq 0$.

The gain in the energy is caused by the considered above Kohn anomaly [14] which implies that for $\mathcal{P} = \mathcal{P}_{CDW}$ (in 1D–metal $\mathcal{P}_{CDW} = 2k_F$, $\omega_0(2k_F) = 0$, where $k_F$ is a Fermi momentum) the phonon frequency $\omega_0(\mathcal{P}_{CDW}, k) = \omega_{CDW}$ softens and, as a consequence, the energy $E_{bp}$ greatly decreases due to a sharp increase in the constant of electron-phonon interaction $\alpha \sim 1/\omega_{CDW}^{1/2}$. For $\omega_{CDW} \to 0$, $\alpha_{CDW} \to \infty$, accordingly $M_{bp} \to \infty$ and CDW appears to be practically immobile.

The general expression for $E_{bp}(\vec{\mathcal{P}})$ is complicated and even in the case of $\vec{\mathcal{P}} = 0$, there are only variational estimates for it [12]. It can be assumed that the general form of the $E_{bp}$ dependence on $\mathcal{P}$ will be δ-shaped, in which the δ-shaped minimum will correspond to $\mathcal{P} = \mathcal{P}_{CDW}$ (Figure 1).

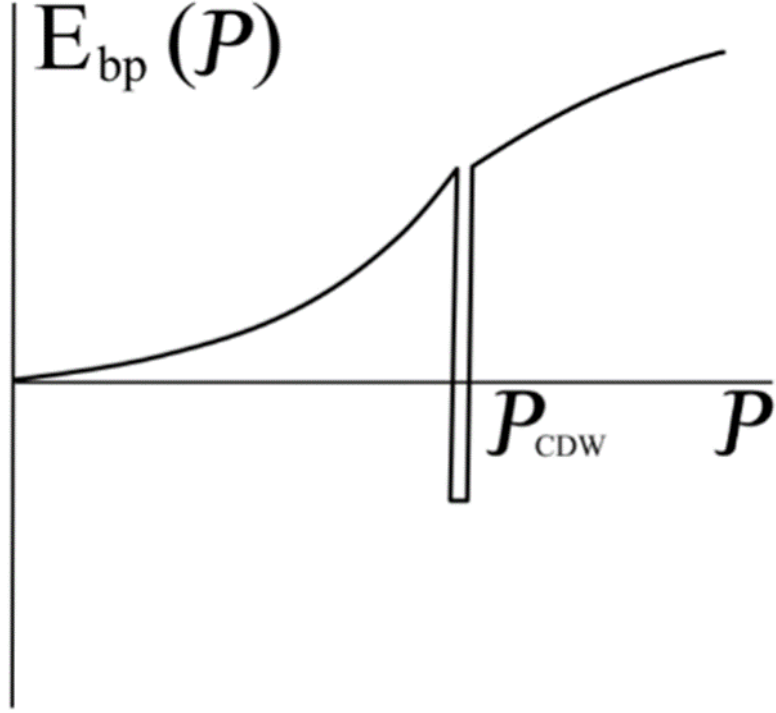

Figure 1. The suggested TI-bipolaron spectrum for charge density wave. It looks as roton spectrum, but is more sharp at $\vec{\mathcal{P}} = \vec{\mathcal{P}}_{CDW}$ due to Kohn anomaly

For this spectrum, TI-bipolarons will pass on to the state with the energy minimum $E_{bp}(\mathcal{P}_{CDW})$, to form a single charge density wave consisting of paired states and determined by expression (4).

Hence, for $T_{CDW}, T_{bp} > T_C$, where $T_{CDW} = |E_{bp}(\mathcal{P}_{CDW}) - E_{bp}(0)|$, $T_{bp} = |E_{bp}(0) - 2E_p(0)|$, $E_p(0)$ - is the polaron ground state energy, the pseudogap precedes the SC. If in this case the condition $T_{CDW} > T_{bp}$ is fulfilled, then the pseudogap is a coherent pseudophase, and for $T_{bp} > T_{CDW}$, a coherent pseudophase is preceded by a non-coherent phase of free pairs. If the inequality $T_{CDW}, T_{bp} > T_C$ is not fulfilled, then the smallest quantity ($T_{CDW}$ or $T_{bp}$) becomes equal to $T_C$. In any case the superconducting phase coexists with the pseudogap one which disappears for $T = 0$, when all the pairs are in a Bose condensate and the CDW amplitude vanishes.

Energetical advantageousness of a condensate phase follows from expressions (7), (8), which suggest that a homogeneous Bose condensate has a lower energy on condition:

$$\mathcal{P} < M_{bp}\sqrt{2\omega_0(\mathcal{P},k)/M} \qquad (9)$$

It should be noted that the scenario considered is close in many respects to the Fröhlich superconductivity model [15]. In the Fröhlich model it was assumed that two electrons with opposite momenta (as in BCS) on the Fermi surface are connected by phonon with the wave vector $\mathcal{P}_{CDW}$ ($2k_F$ in 1D case), to form thereby a charged phonon. Being bosons, such phonons, in a macroscopic number, can be in the same state with the wave vector $\mathcal{P}_{CDW}$ to form a CDW (Fröhlich charge wave). Such a wave, however, will not be superconducting, since pinning which always takes place in real crystals or its scattering by normal carriers will hinder such a wave. The main difference between the TI-bipolaron description and Fröhlich approach is in the formation of a Bose condensate of TI-bipolarons (which form a wave of Fröhlich charged phonons) which is just responsible for superconductivity.

## 3    Discussion – comparison with experiment

To be more specific let us consider the case of a HTSC such as YBCO. The vector of a CDW in YBCO lies in *ab*-plane and has two equally likely directions: along *a*-axis $(\vec{\mathcal{P}}_{CDW,x})$ and along *b*-axis $(\vec{\mathcal{P}}_{CDW,y})$, corresponding to antinodal directions. For $T_c < T < T^*$ these directions of $\vec{\mathcal{P}}_{CDW}$ correspond to nonzero soft phonon modes $\omega_0(\mathcal{P}_{CDW}, k)$. The occurrence of a CDW in these directions in YBCO was detected in a lot of experiments including those on nuclear magnetic resonance [16-18], resonant inelastic X-ray scattering [19-21], resonant scattering and diffraction of hard X-rays [22-23]. Relevant softening of phonon modes in the course of a CDW formation was also observed [24].

Still greater softening of phonon modes can be expected for $T < T_c$. A phonon mode corresponding to a nodal direction can vanish which corresponds to a lack of a gap in the nodal direction. This fact was confirmed experimentally [22] for $Bi_2Sr_{2-x}La_xCuO_{6+\delta}$ with the use of combined methods of resonant X-ray scattering, scanning tunneling microscopy and angle-resolved photoelectron spectroscopy. In the case of YBCO with $\omega_0(\mathcal{P}_{CDW}, k) = \Delta_0|\cos k_x a - \cos k_y a|$ (absolute value of CDW pairing amplitudes, or energy gap [25-26] which are the renormalized phonon frequency in TI-bipolaron theory ) according to (9), any value of $\mathcal{P}_{CDW}$ leads to instability of a CDW and formation of a SC phase in this direction while retaining the pseudogap state in the antinodal direction.

It should be noted that in the approach suggested a difference between CDW and PDW disappears and $\mathcal{P}_{CDW} = \mathcal{P}_{PDW}$ (ref. [27]).

## References

1. Larkin A., Varlamov A. *Theory of fluctuations in superconductors*, Oxford University Press, Oxford UK (2005). DOI: https://doi.org/10.1093/acprof:oso/9780198528159.001.0001

2. D. F. Agterberg, J.C. Séamus Davis, S. D. Edkins, E. Fradkin, D. J. Van Harlingen, S.A. Kivelson, P. A. Lee, L. Radzihovsky, J. M. Tranquada, Y. Wang, The Physics of Pair-Density Waves: Cuprate Superconductors and Beyond. *Annual Review of Condensed Matter Physics* **11**, 231-270 (2020). https://doi.org/10.1146/annurev-conmatphys-031119-050711

3. Grüner G. *Density waves in solids*, Addison-Wesley, Reading (1994). DOI https://doi.org/10.1201/9780429501012

4. V. D. Lakhno. Superconducting Properties of 3D Low-Density Translation-Invariant Bipolaron Gas, *Advances in Condensed Matter Physics*, **2018**, 1380986 (2018). https://doi.org/10.1155/2018/1380986

5. V. D. Lakhno. Superconducting properties of a nonideal bipolaron gas, *Physica C: Superconductivity and its Applications*, **561**, 1-8 (2019). https://doi.org/10.1016/j.physc.2018.10.009

6. V. D. Lakhno. Superconducting Properties of 3D Low-Density TI-Bipolaron Gas in Magnetic Field. *Condens. Matter* **4**(2), 43 (2019). https://doi.org/10.3390/condmat4020043

7. V. D. Lakhno. Translational-Invariant Bipolarons and Superconductivity. *Condens. Matter* **5**(2), 30 (2020). https://doi.org/10.3390/condmat5020030

8 R. Micnas, J. Ranninger, S. Robaszkiewicz, Superconductivity in narrow-band systems with local nonretarded attractive interactions, Rev. Mod. Phys. 62, 113, (1990); DOI: https://doi.org/10.1103/RevModPhys.62.113

9. J. Bardeen, L. N. Cooper, J. R. Schrieffer. Theory of Superconductivity. *Phys. Rev*. **108**, 1175 (1957) DOI: https://doi.org/10.1103/PhysRev.108.1175

10. L.P. Gor'kov. Microscopic Derivation of the Ginzburg-Landau Equations in the Theory of Superconductivity. *Sov. Phys. JETP* **9**, 1364 (1959). http://www.jetp.ras.ru/cgi-bin/e/index/e/9/6/p1364?a=list

11. Grimvall G. *The Electron-Phonon Interaction in Metals*, North-Holland Publ. Comp., Amsterdam, (1981) ISBN: 9780444861054

12. V. D. Lakhno. Pekar's ansatz and the strong coupling problem in polaron theory. *Phys.-Usp*. **58** 295 (2015). https://doi.org/10.3367/UFNe.0185.201503d.0317

13 T. Lee, F. Low, D. Pines, The motion of slow electrons in a polar crystal, Phys. Rev. 90, 297, (1953) : https://doi.org/10.1103/PhysRev.90.297

14. W. Kohn. Image of the Fermi Surface in the Vibration Spectrum of a Metal. *Phys. Rev. Lett*. **2**, 393 (1959). DOI: https://doi.org/10.1103/PhysRevLett.2.393

15. H. Fröhlich. On the theory of superconductivity: the one-dimensional case. *Proc. R. Soc. Lond. A* **223**, 296–305 (1954). https://doi.org/10.1098/rspa.1954.0116

16. T. Wu, H. Mayaffre, S. Krämer, M. Horvatić, C. Berthier, W. N. Hardy, R. Liang, D. A. Bonn, M.-H. Julien. Magnetic-field-induced charge-stripe order in the high-temperature superconductor $YBa_2Cu_3O_y$. *Nature* **477**, 191–194 (2011). https://doi.org/10.1038/nature10345

17. T. Wu, H. Mayaffre, S. Krämer, M. Horvatić, C. Berthier, P. L. Kuhns, A. P. Reyes, R. Liang, W. N. Hardy, D. A. Bonn, M.-H. Julien. Emergence of charge order from the vortex state of a high-temperature superconductor. *Nat Commun* **4**, 2113 (2013). https://doi.org/10.1038/ncomms3113

18. T. Wu, H. Mayaffre, S. Krämer, M. Horvatić, C. Berthier, W.N. Hardy, R Liang, D.A. Bonn, M.-H. Julien. Incipient charge order observed by NMR in the normal state of $YBa_2Cu_3O_y$. *Nat Commun* **6**, 6438 (2015). https://doi.org/10.1038/ncomms7438

19. G. Ghiringhelli, M. Le Tacon, M. Minola, S. Blanco-Canosa, C. Mazzoli, N. B. Brookes, G. M. De Luca, A. Frano, D. G. Hawthorn, F. He, T. Loew, M. Moretti Sala, D. C. Peets, M. Salluzzo, E. Schierle, R. Sutarto, G. A. Sawatzky, E. Weschke, B. Keimer, L. Braicovich. Long-Range Incommensurate Charge Fluctuations in $(Y,Nd)Ba_2Cu_3O_{6+x}$. *Science* **337**(6096), 821-825 (2012) DOI: https://doi.org/10.1126/science.1223532

20. A. J. Achkar, R. Sutarto, X. Mao, F. He, A. Frano, S. Blanco-Canosa, M. Le Tacon, G. Ghiringhelli, L. Braicovich, M. Minola, M. Moretti Sala, C. Mazzoli, Ruixing Liang, D. A. Bonn, W. N. Hardy, B. Keimer, G. A. Sawatzky, D. G. Hawthorn. Distinct Charge Orders in the Planes and Chains of Ortho-III-Ordered $YBa_2Cu_3O_{6+\delta}$ Superconductors Identified by Resonant Elastic X-ray Scattering. *Phys. Rev. Lett*. **109**, 167001 (2012). DOI: https://doi.org/10.1103/PhysRevLett.109.167001

21. E. Blackburn, J. Chang, M. Hücker, A. T. Holmes, N. B. Christensen, Ruixing Liang, D. A. Bonn, W. N. Hardy, U. Rütt, O. Gutowski, M. v. Zimmermann, E. M. Forgan, S. M. Hayden. X-Ray Diffraction Observations of a Charge-Density-Wave Order in Superconducting Ortho-II YBa$_2$Cu$_3$O$_{6.54}$ Single Crystals in Zero Magnetic Field. *Phys. Rev. Lett*. **110**, 137004 (2013). DOI: https://doi.org/10.1103/PhysRevLett.110.137004

22. R. Comin, A. Frano, M. M. Yee, Y. Yoshida, H. Eisaki, E. Schierle, E. Weschke, R. Sutarto, F. He, A. Soumyanarayanan, Yang He, M. Le Tacon, I. S. Elfimov, Jennifer E. Hoffman, G. A. Sawatzky, B. Keimer, A. Damascelli. Charge Order Driven by Fermi-Arc Instability in Bi$_2$Sr$_{2-x}$La$_x$CuO$_{6+\delta}$. *Science* **343**(6169), 390-392 (2014). DOI: https://doi.org/10.1126/science.1242996

23. J. Chang, E. Blackburn, A. T. Holmes, N. B. Christensen, J. Larsen, J. Mesot, Ruixing Liang, D. A. Bonn, W. N. Hardy, A. Watenphul, M. v. Zimmermann, E. M. Forgan, S. M. Hayden. Direct observation of competition between superconductivity and charge density wave order in YBa$_2$Cu$_3$O$_{6.67}$. *Nature Phys* **8**, 871–876 (2012). https://doi.org/10.1038/nphys2456

24. M. Le Tacon, A. Bosak, S. M. Souliou, G. Dellea, T. Loew, R. Heid, K-P. Bohnen, G. Ghiringhelli, M. Krisch, B. Keimer. Inelastic X-ray scattering in YBa$_2$Cu$_3$O$_{6.6}$ reveals giant phonon anomalies and elastic central peak due to charge-density-wave formation. *Nature Phys* **10**, 52–58 (2014). https://doi.org/10.1038/nphys2805

25. R. Comin, R. Sutarto, F. He, E. H. da Silva Neto, L. Chauviere, A. Fraño, R. Liang, W. N. Hardy, D. A. Bonn, Y. Yoshida, H. Eisaki, A. J. Achkar, D. G. Hawthorn, B. Keimer, G. A. Sawatzky, A. Damascelli, Symmetry of charge order in cuprates. *Nature Materials* **14**, 796–800 (2015). https://doi.org/10.1038/nmat4295

26. D. Chowdhury, S. Sachdev, Density-wave instabilities of fractionalized Fermi liquids. *Phys. Rev. B* **90**, 245136 (2014) DOI: https://doi.org/10.1103/PhysRevB.90.245136

27. M. H. Hamidian, S. D. Edkins, Sang Hyun Joo, A. Kostin, H. Eisaki, S. Uchida, M. J. Lawler, E.-A. Kim, A. P. Mackenzie, K. Fujita, Jinho Lee, J. C. Séamus Davis. Detection of a Cooper-pair density wave in Bi$_2$Sr$_2$CaCu$_2$O$_{8+x}$. *Nature* **532**, 343–347 (2016). https://doi.org/10.1038/nature17411